# Light Weight CNN for classification of Brain Tumors from MRI Images

Natnael Alemayehu
Computer Science *department; Georgia State University*

*Abstract*— This study presents a convolutional neural network (CNN)-based approach for the multi-class classification of brain tumors using magnetic resonance imaging (MRI) scans. We utilize a publicly available dataset containing MRI images categorized into four classes: glioma, meningioma, pituitary tumor, and no tumor. Our primary objective is to build a light weight deep learning model that can automatically classify brain tumor types with high accuracy. To achieve this goal, we incorporate image preprocessing steps, including normalization, data augmentation, and a cropping technique designed to reduce background noise and emphasize relevant regions. The CNN architecture is optimized through hyperparameter tuning using Keras Tuner, enabling systematic exploration of network parameters. To ensure the robustness and generalizability of the model, we apply 5-fold cross-validation, where each hyperparameter configuration is evaluated across multiple data splits to mitigate overfitting. Experimental results demonstrate that the proposed model achieves a classification accuracy of 98.78%, indicating its potential as a diagnostic aid in clinical settings. The proposed method offers a low-complexity yet effective solution for assisting in early brain tumor diagnosis.

## I. INTRODUCTION

Brain tumors are among the most severe and life-threatening neurological disorders, often requiring timely diagnosis and intervention. Magnetic Resonance Imaging (MRI) is widely used in clinical practice due to its ability to produce high-resolution images of soft tissues. However, analyzing MRI scans manually is a time-consuming and subjective task that relies heavily on radiologist expertise. As a result, there is growing interest in automated approaches that can assist clinicians by accurately classifying brain tumors using computational models [1].

Deep learning techniques particularly Convolutional Neural Networks (CNNs) have demonstrated remarkable success in various image classification tasks, including medical imaging [2]. Many existing approaches for brain tumor classification rely on large pre-trained CNN architectures such as VGG, ResNet, or Inception, which, while accurate, are often computationally expensive and memory-intensive.

In this study, we demonstrate that a custom-designed, lightweight CNN can achieve accuracy scores comparable to those of deeper, pre-trained networks, without incurring the same computational costs. Our model is specifically tailored for brain tumor classification across four categories: glioma, meningioma, pituitary tumor, and no tumor. To enhance performance and generalization, we apply preprocessing techniques such as rescaling, data augmentation, and a contour-based cropping strategy that removes irrelevant background from MRI scans. We also integrate Keras Tuner to optimize both architectural and training hyperparameters systematically. Furthermore, we employ 5-fold cross-validation to ensure robust evaluation and minimize the risk of overfitting to a single data split.

The rest of this paper is organized as follows: Section II reviews related work in the area of brain tumor classification using CNNs. Section III describes the dataset, preprocessing steps, and model design. Section IV presents experimental setup and evaluation results. Section V concludes the paper and suggests directions for future work.

The code and resources used in this study are available at: **GitHub Link**

## II. RELATED WORKS

Convolutional Neural Networks (CNNs) have emerged as the leading approach for brain tumor classification, owing to their capacity for automatic feature extraction and good performance in medical image analysis.

Talukder et al [3] proposed a transfer learning framework that integrates extensive preprocessing, architecture reconstruction, and fine-tuning. They evaluated four deep pre-trained models—Xception, ResNet50V2, InceptionResNetV2, and DenseNet201—on the Figshare brain MRI dataset, which consists of 3,064 images. Among these, ResNet50V2 achieved the highest classification accuracy of 99.68%, demonstrating the efficacy of fine-tuned architectures in extracting discriminative features.

Deepak and Ameer [4] adopted a hybrid approach by leveraging GoogLeNet for deep feature extraction and combining it with conventional classifiers. Their method, evaluated using patient-level five-fold cross-validation on the Figshare dataset, achieved a mean classification accuracy of 98%, highlighting the utility of transfer learning when labeled medical data is limited.

Ismael et al. [5] introduced a classification model based on Residual Networks (ResNet), optimized for brain tumor MRI

classification. Their model achieved an accuracy of 99% on the same Figshare dataset. Residual connections improved gradient flow and enabled deeper network design, resulting in stronger representation learning.

Afshar et al. [6] explored Capsule Networks (CapsNets) as an alternative to CNNs, capitalizing on their ability to preserve spatial hierarchies and handle affine transformations more effectively. Their work investigated the overfitting behavior of CapsNets and their performance on both full-brain and segmented tumor images. The proposed method outperformed conventional CNNs, particularly when training data was limited.

In contrast to these approaches, the present work proposes a lightweight, custom CNN architecture trained from scratch. Rather than relying on pre-trained models, our method incorporates preprocessing, data augmentation, and hyperparameter optimization using Keras Tuner.

### III. MATERIALS AND METHODS
#### A. Data Description and Characterization

The dataset used in this study is a curated combination of three publicly available brain MRI datasets on: Figshare, SARTAJ, and Br35H [7]. It comprises a total of 7,023 MRI images, categorized into four classes: glioma, meningioma, pituitary tumor, and no tumor. The "no tumor" class images were sourced specifically from the Br35H dataset. Out of the 7023 Images 5712 were categorized as Training data and 1311 were categorized as Test data. Below we visualize the distribution of the various classes in each data category.

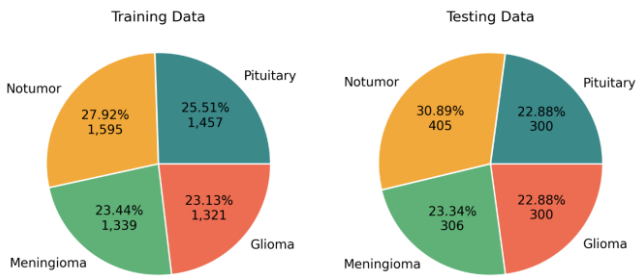

Fig. 1. Tumor Type Distribution in Training and Testing Data.

All images are grayscale brain MRIs provided in JPEG format and organized into subdirectories according to their class labels. The dataset contains images of varying sizes and resolutions, requiring preprocessing to standardize input for the convolutional neural network.

#### B. Image Preprocessing

The dataset contained images that had irrelevant background regions. To reduce spatial redundancy and eliminate irrelevant background, a contour-based cropping step was applied as part of the preprocessing pipeline. This method aimed to isolate the brain region from surrounding non-informative areas in the original MRI images, such as black borders and scanner artifacts. Each image was first converted to grayscale, then smoothed using Gaussian blurring to reduce high-frequency noise. Binary thresholding was applied to separate foreground from background, followed by morphological operations (erosion and dilation) to remove small, spurious components and close gaps within the brain region. Contours were extracted from the cleaned binary mask using edge-following algorithms. The largest external contour was presumed to correspond to the brain structure. From this contour, the extreme points (leftmost, rightmost, topmost, and bottommost) were computed to define a tight bounding box. The original image was then cropped according to this bounding box, effectively isolating the brain tissue and discarding surrounding irrelevant regions.

The cropped image was then resized to a fixed input shape of 150×150×3 to ensure consistency during training. This preprocessing step was applied uniformly to all images in the dataset and was not evaluated as an isolated variable. Its primary goal was to simplify the input and allow the model to focus on the tumor-bearing brain region, potentially improving convergence and classification accuracy by removing irrelevant visual information.

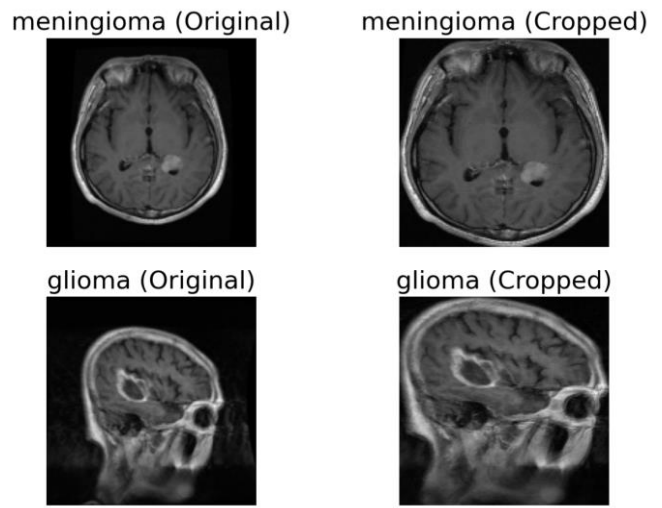

Fig. 2. Example of original and cropped brain MRI images for two tumor classes.

To mitigate overfitting and enhance generalization on a limited dataset, a range of data augmentation techniques was applied during model training. These augmentations were implemented using Keras's ImageDataGenerator and included transformations designed to simulate natural variations in MRI acquisition without altering the diagnostic content of the images. Specifically, we applied small rotations (up to ±10°) to simulate head movement, brightness adjustments (±15%) to account for scanner intensity variations, shear transformations (up to 12.5%) to mimic minor distortions, horizontal flipping to simulate mirrored anatomical presentations, and minor translations (up to 0.2% shift) in both height and width. Furthermore, pixel values were normalized to a [0, 1] scale by dividing each value by 255 in order to help stabilize training.

All augmentations were performed in real-time during training, such that each epoch presented the model with slightly different versions of the same images. This strategy effectively increased the diversity of training data and reduced reliance on memorizing specific spatial patterns, thus promoting more diversified feature learning across the dataset.

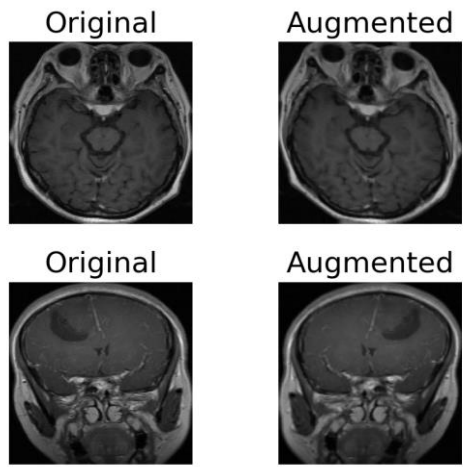

Fig. 3. Examples of orginal vs augmented brain MRI images.

### C. Model Architecture

The proposed custom CNN architecture begins with an input layer that accepts RGB images of size 150 × 150×3 pixels. This is followed by a sequence of four convolutional layers, each employing ReLU activation functions and 'same' padding to preserve spatial dimensions.

Specifically, the convolutional layers are configured with 32, 128, 128, and 128 filters, respectively, using kernel sizes of 3 × 3, 4 × 4, 3 × 3, and 3 × 3. This design enables the network to capture hierarchical and increasingly abstract spatial features. After the convolutional blocks, the feature maps are flattened and passed through a fully connected dense layer comprising 384 neurons with ReLU activation. A dropout layer with a rate of 0.3 is applied subsequently to mitigate overfitting by randomly deactivating neurons during training.

The final output layer consists of four neurons activated by a softmax function, corresponding to the four target classes: glioma, meningioma, pituitary tumor, and no tumor.

To improve training stability and convergence, two callbacks were employed: EarlyStopping, configured with a patience of 8 and a minimum delta of 1e-4, to terminate training when the validation loss ceased improving; and ReduceLROnPlateau, configured with a patience of 5, a reduction factor of 0.3, and a minimum learning rate of 1e-6, to dynamically lower the learning rate when validation performance stagnated.

The model was compiled using the Adamax optimizer with a learning rate of $9.534 \times 10^{-4}$ and optimized with the categorical cross-entropy loss function, monitoring accuracy as the evaluation metric.

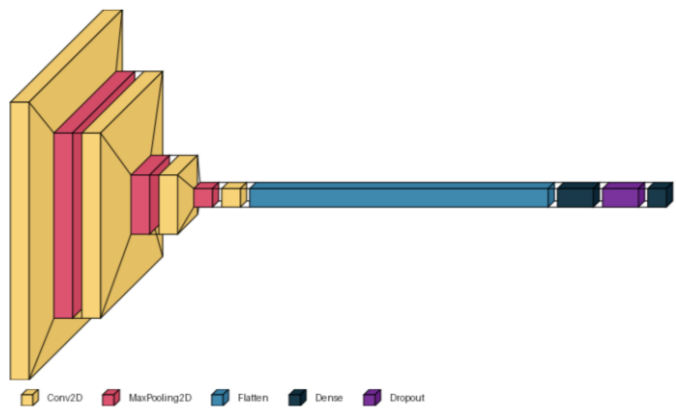

Fig. 4. Proposed CNN architecture used for brain tumor classification.

The model was trained on 80% of the training data set and validated on the remaining 20% of the training data set. Each training iteration used a batch size of 32 and was conducted over 50 epochs with real-time data augmentation applied via ImageDataGenerator.

To further optimize model performance, we employed Keras Tuner for hyperparameter search, integrating it with k-fold cross-validation to ensure more reliable hyperparameter evaluation and selection. A custom CV-Tuner class was implemented by extending the standard RandomSearch tuner to perform 5-fold cross-validation within each trial, thus producing more reliable estimates of each hyperparameter configuration's performance. The hyperparameter search space included the number of filters for each convolutional layer (chosen among 32, 64, and 128), kernel sizes (3 × 3 or 4 × 4), the number of dense layer units (ranging from 256 to 512 in increments of 64), dropout rates (ranging from 0.3 to 0.6), and learning rates (sampled logarithmically between 1e-4 and 1e-2).

Each hyperparameter configuration was evaluated using full 5-fold cross-validation, with a maximum of four trials (max_trials=4). During each fold, the training data was augmented using random rotations, brightness adjustments, slight translations, shearing, and horizontal flips to improve model generalization. EarlyStopping and ReduceLROnPlateau callbacks were applied in each fold to prevent overfitting and dynamically adjust the learning rate, respectively.

Finally, the selected hyperparameters were used to retrain a final model from scratch using the entire preprocessed training dataset. The model's performance was subsequently evaluated on an independent, unseen test set to assess its generalization capability.

D. Evaluation and Interpretation

To comprehensively assess the performance of the proposed CNN architecture, we adopted a two-tiered evaluation framework: 5-fold stratified cross-validation during hyperparameter optimization and independent testing on a held-out dataset. This strategy allowed for both internal validation during model tuning and external validation on unseen data, reducing the risk of overfitting and providing a more generalizable performance estimate.

The evaluation metrics used include:

**1) Accuracy**

The ratio of correctly classified samples to the total number of samples:

$$\text{Accuracy} = \frac{\sum_{i=1}^{C} TP_i}{\sum_{i=1}^{C}(TP_i + FP_i + FN_i + TN_i)}$$

where $TP_i, FP_i, FN_i$, and $TN_i$ represent the true positives, false positives, false negatives, and true negatives for class i, and C is the total number of classes.

**2) Precision**

The proportion of positive identifications that were actually correct:

$$\text{Recall}_i = \frac{TP_i}{TP_i + FN_i}$$

**3) Recall (Sensitivity)**

The proportion of actual positives that were correctly identified:

$$\text{Recall}_i = \frac{TP_i}{TP_i + FN_i}$$

**4) F1-Score**

The harmonic mean of precision and recall:

$$F1_i = 2 \cdot \frac{\text{Precision}_i \cdot \text{Recall}_i}{\text{Precision}_i + \text{Recall}_i}$$

These metrics were calculated per class and summarized using Macro Average which is the unweighted mean across all classes and Weighted Average which is the mean weighted by class support (number of true instances per class).

This multi-metric evaluation is particularly important in medical imaging tasks, where certain tumor classes may be underrepresented and a high overall accuracy can obscure poor performance on rare or critical classes.

Furthermore, final test set predictions were analyzed via a classification report and confusion matrix, enabling detailed per-class error analysis.

IV. RESULTS

The model achieved consistently strong results across the various evaluation metrics.

The initial proposed model achieved a test accuracy of 98.24%. Following hyperparameter optimization using Keras Tuner integrated with 5-fold cross-validation, the model was retrained on the entire training set and evaluated on an independent test set. The final model achieved a test accuracy of 98.78%, demonstrating a slightly improved generalization capability.

The best hyperparameter configuration identified was as follows: 32 filters with a $4 \times 4$ kernel in the first convolutional layer, 64 filters with a $3 \times 3$ kernel in the second layer, and 128 filters with a $3 \times 3$ kernel in the third convolutional layer, followed by 128 filters with a $4 \times 4$ kernel in the fourth convolutional layer. The dense layer comprised 512 units, with a dropout rate of 0.5 applied before the output layer. The optimal learning rate was found to be $1.19 \times 10^{-3}$. This configuration resulted the model to have 4,150,766 total parameters which is far less than the pre-trained models that typically have over 10 Million parameters.

TABLE I. CLASSIFICATION REPORT

| Class | Precision | Recall | F1-score | Support |
|---|---|---|---|---|
| Giloma | 0.99 | 0.99 | 0.99 | 300 |
| Meningioma | 0.99 | 0.97 | 0.98 | 306 |
| No Tumor | 0.99 | 1.00 | 0.99 | 405 |
| Pituitary | 0.98 | 0.99 | 0.99 | 300 |
| **Accuracy** | | | 0.99 | 1311 |
| Macro Avg | 0.99 | 0.99 | 0.99 | 1311 |
| Weighted Avg | 0.99 | 0.99 | 0.99 | 1311 |

These results suggest that the model has high precision score across all four tumor categories. It is also noted that the model had a lower recall value for Meningioma Class than other classes.

In addition to the classification metrics, a confusion matrix is plotted to visually examine misclassification patterns (Fig. 5). While the classification report quantifies class-wise performance, the matrix reveals specific error trends. For example, *Glioma* samples were occasionally misclassified as *Pituitary*, likely due to overlapping radiological characteristics in certain slices. On the other hand, the *Notumor* class achieved perfect classification, with all 405 samples correctly identified. This further validates the model's ability to detect the absence of tumor-related features. It is also noted that Meningioma was wrongly classified as No tumor in some instances, which is not desirable. The matrix also confirms that false positives and false negatives are relatively low across all classes, supporting the high F1-scores observed earlier.

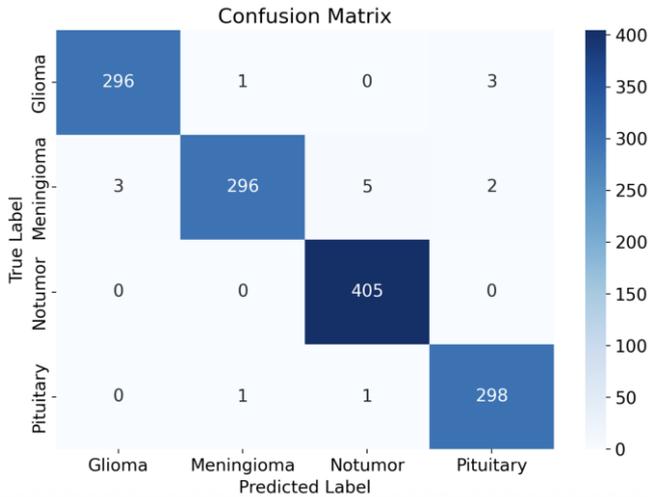

Fig. 5. Confusion matrix of the CNN model evaluated on the test dataset

To gain insights into the internal representations learned by the CNN, we visualized the feature maps produced after the first and second convolutional layers for a simple Glioma Type tumor image. As shown in Figure 6, the initial layer captures fundamental low-level features such as edges and contours, which define the brain's boundary. By the second convolutional layer, the model begins to extract more abstract patterns, including textural differences and central anatomical structures. This progressive feature extraction confirms that the network is learning hierarchical representations relevant for distinguishing tumor types. Such interpretability strengthens the clinical applicability of the proposed approach.

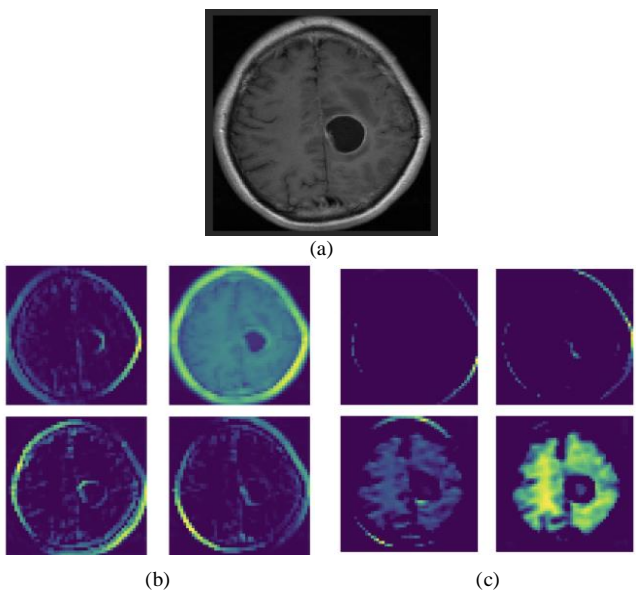

Fig. 6. Visualization of learned features of a sample Glioma Class Tumor. (a) Original input image. (b) Feature maps from the first convolutional layer. (c) Feature maps from the second convolutional layer.

## V. DISCUSSION AND CONCLUSION

The experimental findings demonstrate that the proposed convolutional neural network (CNN) model achieved high performance in the task of multi-class brain tumor classification. With a test accuracy of 98.78% and strong class-wise precision, recall, and F1-scores, the model exceeded initial expectations. These results were obtained using a relatively lightweight architecture and limited dataset size, without the need for large-scale pre-trained networks, indicating that a well-designed and well-tuned CNN can be both efficient and highly effective in medical image classification tasks.

Several design choices contributed to the model's success. Data augmentation techniques, including rotation, brightness variation, and minor shifts, helped improve generalization by exposing the model to a broader range of input conditions. Hyperparameter tuning via Keras Tuner, combined with 5-fold cross-validation, further ensured that the architecture was well-calibrated and generalized effectively across data splits.

However, the study is not without limitations. Due to time and computational constraints associated with k-fold cross-validation, the hyperparameter optimization process was limited to only four search trials using Keras Tuner. A larger number of trials may have led to further improvements in model performance. Additionally, the model was trained and evaluated on a single imaging modality, MRI jpg format images, and on a dataset restricted to four distinct tumor categories. As such, the generalizability of the model to more complex or heterogeneous datasets, including multi-modal MRI scans or cases involving tumor progression, remains unverified.

Future work may involve extending the pipeline to support additional imaging modalities, increasing the trails for hyperparameter optimization, and increasing data size. Despite these limitations, the present work demonstrates that a lightweight, custom CNN when properly optimized and trained can deliver high accuracy and efficiency, making it suitable for real-world diagnostic applications, particularly in resource-constrained environments.

ACKNOWLEDGMENT

The author would like to thank the contributors of the brain tumor MRI dataset published on Kaggle, which was compiled from the Figshare, SARTAJ, and Br35H datasets. Their efforts in collecting and organizing the data made this research possible. This work was conducted as part of an academic project and did not receive external funding.